\documentclass[conference, compsocconf]{IEEEtran}
\usepackage[pdftex]{graphicx}
\usepackage{epstopdf}
\usepackage{rotating}
\usepackage{amsmath}
\usepackage{amssymb}
\usepackage{graphicx}
\usepackage{algorithm}
\usepackage{algorithmic}
\usepackage{subfigure}
\usepackage{threeparttable}

\usepackage{amsmath}
\usepackage{amssymb}

\newtheorem{defn}{Definition}

\newtheorem{problem}{Problem}

\begin{document}
%
\title{Granular association rules for multi-valued data}


\author{\IEEEauthorblockN{Fan Min and William Zhu}
\IEEEauthorblockA{Lab of Granular Computing, Zhangzhou Normal University, Zhangzhou 363000, China.\\
Email: minfanphd@163.com, williamfengzhu@gmail.com
}}
\maketitle

\begin{abstract}
Granular association rule is a new approach to reveal patterns hide in many-to-many relationships of relational databases.
Different types of data such as nominal, numeric and multi-valued ones should be dealt with in the process of rule mining.
In this paper, we study multi-valued data and develop techniques to filter out strong however uninteresting rules.
An example of such rule might be ``male students rate movies released in 1990s that are \emph{not} thriller."
This kind of rules, called negative granular association rules, often overwhelms positive ones which are more useful.
To address this issue, we filter out negative granules such as ``\emph{not} thriller" in the process of granule generation.
In this way, only positive granular association rules are generated and strong ones are mined.
Experimental results on the movielens data set indicate that most rules are negative, and our technique is effective to filter them out.
\end{abstract}


  %
  %
  \section{Introduction}\label{section: introduction}
Granular association rule \cite{MinHuZhu12GranularTwo,MinHuZhu12GranularFour} is a new approach to build recommender systems \cite{BalabanovicM1997Fab,Goldberg1992usingcollaborative,Yao95Measuring}.
The data model is a many-to-many entity-relationship system (MMER) which is composed of two information systems and a relation between them \cite{MinHuZhu12GranularFour}.
For example, the movielens data set \cite{movielens} is composed of users, movies, and the rating of movies by users.
Suppose we are interested in what kind of users rate what kind of movies.
``Women rate horror movies" and ``male students rate thriller movies released in 1995" might be two interesting granular association rules.
Here we observe that both sides of a granular association rule can take different number of attributes, therefore they have different granules \cite{Lin98Granular,YaoYao03Information,Yao00Granular,Zadeh97Towards,ZhuWang03Reduction}.
This is the major difference of this types of rules from other relational association rules (e.g., \cite{AftratiF2012Chains,DehapseL1998Finding,Goethals2010Mining}).

The original definition of granular association rule \cite{MinHuZhu12GranularFour} considers only nominal data.
In applications, there are other types such as numeric, multi-valued, and interval valued data.
Numeric data might be the most important type in applications.
For example, in the movielens data set, each movie has a release date.
It is hard to construct strong rules using this information directly since few movies are released in the same day.
We would like to use release year, or even the release decade instead to obtain coarser granules and stronger rules.

In this paper, we consider multi-valued data which are also common in applications.
In the movielens data set, each movie may belong 1 to 18 genres, including \texttt{action}, \texttt{adventure}, \texttt{children}, and so on.
However, multi-valued data cannot be stored directly into relational databases.
We need to preprocess them before constructing information systems and MMERs.
There are at least three approaches to this issue.
\begin{enumerate}
\item{
Combine existing movie genres to form new ones.
For example, \texttt{action + children} is a new genre.
In this way, if a movie is \texttt{action + children}, it is neither \texttt{action} nor \texttt{children}.
Hence this approach is unreasonable from the semantic viewpoint.}
\item{
Assign a priority to each genre and keep only the most important one for a movie \cite{MinZhu12Parametric}.
For example, if a movie is \texttt{action + children}, we will view it only as \texttt{action}.
The drawbacks are also obvious: many interesting rules cannot be found.}
\item{
Scale the movie genre attribute into 18 boolean attributes.
With this approach, we obtain ``male students rate movies released in 1990s that are \emph{not} thriller," which is stronger than ``male students rate thriller movies released in 1990s."
This is because that each user rate only a small fraction of all movies.
Therefore we need to filter out this kind of uninteresting rules.}
\end{enumerate}

We adopt the third approach and amend the drawback directly.
For this purpose we define positive granules, positive granular association rules and negative ones.
A granule is positive if and only if all attribute values of the scaled data are true.
For example, ``thriller movies released in 1990s" is a positive granule, while ``movies released in 1990s that are \emph{not} thriller" is a negative granule.
A granular association rule is positive if and only if both sides of the rule are positive granules.
For brevity, a positive (negative) granular association rule will be called a positive (negative) rule.

We propose an algorithm with three main steps to mine all strong positive rules satisfying thresholds of four measures \cite{MinHuZhu12GranularFour}.
Step 1, generate positive granules with length one.
Step 2, produce longer positive granules following the structure of the Apriori algorithm \cite{AgrawalR1994Apriori}.
Naturally, only positive granules satisfying coverage measures are kept.
Step 3, generate candidate rules through connecting positive granules on two universes, and check wether or not these rules satisfy the confidence thresholds.
A technique developed in \cite{MinZhu12ParametricCold} is employed to speed up the third step.

Experiments are undertaken on the movielens data set \cite{movielens}.
We have a number of observations.
First, many interesting rules are lost if we adopt the priority assigning approach.
Therefore the priority-based approach is unacceptable.
Second, if we do not filter out negative granular association rules, they will overwhelm positive ones.
In fact, with thresholds settings that generates thousands of rules, not even one positive rule is generated.
In summary, our algorithm keeps all interesting rules, and at the same time filters out a large number of uninteresting rules.

  %
  %
  \section{Positive rules}\label{section: granular-association-rules}
In this section, we define positive granules and positive rules.
Since granules and granular association rules have been well defined in \cite{MinHuZhu12GranularFour}, we will focus on new ones.

  %
  %
  \subsection{Information systems and granules}\label{subsection: ins-granules}
The data model is based on information systems and binary relations.
\begin{defn}\label{definition: ins}
$S = (U, A)$ is an information system, where $U = \{x_1, x_2, \dots, x_n\}$ is the set of all objects, $A = \{a_1, a_2, \dots, a_m\}$ is the set of all attributes, and $a_j(x_i)$ is the value of $x_i$ on attribute $a_j$ for $i \in [1..n]$ and $j \in [1..m]$.
\end{defn}

User information of the movielens data set are stored in an information system given by Table \ref{table: mmer}(a), where $|U| = 943$ and $A$ = \{User-id, Age, Gender, Occupation\}.
This table is different from its original version in two aspects.
First, the Zip-code attribute is removed since they are not useful in the mining process.
Second, the age of the user is discretized according to given intervals $[0, 17]$, $[18, 24]$, \dots, $[56, \infty)$.
In this way, all attributes in Table \ref{table: mmer}(a) are nominal.

In an information system, any $A' \subseteq A$ induces an equivalent relation \cite{Pawlak82Rough,SkowronStepaniuk94Approximation}
\begin{equation}\label{equation: equivalent-relation}
E_{A'} = \{(x, y) \in U \times U| \forall a \in A', a(x) = a(y)\},
\end{equation}
and partitions $U$ into a number of disjoint subsets called \emph{blocks}.
The block containing $x \in U$ is
\begin{equation}\label{equation: block-contain-x}
E_{A'}(x) = \{y \in U| \forall a \in A', a(y) = a(x)\}.
\end{equation}

\begin{defn}\label{defn: granule}
\cite{YaoDeng2013Paradigm} A granule is a triple
\begin{equation}\label{equation: granule}
G = (g, i(g), e(g)),
\end{equation}
where $g$ is the name assigned to the granule, $i(g)$ is a representation of the granule,
and $e(g)$ is a set of objects that are instances of the granule.
\end{defn}

$g = g(A', x)$ is a natural name to the granule.
\begin{equation}\label{equation: intension-granule}
i(g(A', x)) =  \bigwedge_{a \in A'}\langle a: a(x) \rangle.
\end{equation}
\begin{equation}\label{equation: extension-granule}
e(g(A', x)) = E_{A'}(x).
\end{equation}

The \emph{support} of $g(A', x)$ is
\begin{equation}\label{equation: support-granule}
supp(g(A', x)) = supp(\bigwedge_{a \in A'}\langle a: a(x) \rangle) = \frac{|E_{A'}(x)|}{|U|}.
\end{equation}

  %
  %
  \subsection{Scaled attributes and positive granules}\label{subsection: scaled-positive}
A multi-valued attribute has a domain of a power set.
In the movielens data set, there are 18 genres including \texttt{action}, \texttt{children}, \texttt{adventure}, etc.
Since movies can be in several genres at once, the domain of genre is $2^{18}$ instead of 18.
Attribute values include \texttt{\{action\}}, \texttt{\{children\}}, \texttt{\{adventure\}}, \texttt{\{action, children\}}, \texttt{\{action, adventure\}}, etc.
\texttt{unknown} correspond to $\emptyset$.
This attribute can be replaced by 18 boolean attributes, with each indicating whether or not the movie is in the respective genre.
This technique is called scaling \cite{MinLiuFang08Rough} and serves as the foundation of formal concept analysis \cite{GanterB1997Formal}.
In fact, the original data set contain the scaled information instead of the multi-valued one.
It is given by Table \ref{table: mmer}(a).
Here we use release decade instead of release date to obtain a finer granule.

In order to describe this kind of data, we propose the following definition.
\begin{defn}\label{definition: scaled}
Let $S = (U, A)$ be an information system. Any $a \subseteq A$ is a scaled attribute if $a(x) \in \{0, 1\}$, $a(x) = 1$ indicate that $x$ has the attribute specified by $a$, and $a(x) = 0$ for otherwise.
\end{defn}

With scaled attributes identified by the expert, we can focus on granules that are interesting to us.
\begin{defn}\label{definition: interesting-granule}
Let $S = (U, A)$ be an information system and $A_b$ be the set of all scaled attributes. $C = (A', x)$ where $x \in U$, $A' \subseteq A$ is called a \emph{positive granule} iff $\forall a \in A' \cap A_b$, $a(x) \equiv 1$.
\end{defn}
In other words, a positive granule requires that all scaled attributes take true values.
With positive granules identified, we can filter out unimportant granule from the very beginning.

  %
  %
  \subsection{Many-to-many entity-relationship systems}\label{subsection: scaled-positive}
\begin{defn}\label{definition: binary-relation}
Let $U = \{x_1, x_2, \dots, x_n\}$ and $V = \{y_1, y_2, \dots, y_k\}$ be two sets of objects.
Any $R \subseteq U \times V$ is a binary relation from $U$ to $V$.
The neighborhood of $x \in U$ is
\begin{equation}\label{equation: relation}
R(x) = \{y \in V | (x, y) \in R\}.
\end{equation}
\end{defn}

When $U = V$ and $R$ is an equivalence relation, $R(x)$ is the equivalence class containing $x$.
From this definition we know immediately that for $y \in V$,
\begin{equation}\label{equation: relation-reverse}
R^{-1}(y) = \{x \in U | (x, y) \in R\}.
\end{equation}

An example of binary relation is given by Table \ref{table: mmer}(c), where $U$ is the set of users as indicated by Table \ref{table: mmer}(a), and $V$ is the set of movies as indicated by Table \ref{table: mmer}(b).

\begin{table}[tb]\caption{A many-to-many entity-relationship system}\label{table: mmer}
\centering
\setlength{\tabcolsep}{15pt}
\subtable[User]{
\begin{tabular}{cccccc}
\hline
User-id     & Age            & Gender &  Occupation \\
\hline
1           & $[18, 24]$     & M      &  technician \\
2           & $[50, 55]$     & F      &  other      \\
3           & $[18, 24]$     & M      &  writer     \\
\dots       \\
943         & $[18, 24]$     & M      &  student    \\
\hline
\end{tabular}
\label{subtable: user}
}
\qquad
\setlength{\tabcolsep}{2pt}
\subtable[Movie]{
\begin{tabular}{cccccccccc}
\hline
Movie-id       & Release-decade &  Action & Adventure & Animation & \dots & Western \\
\hline
1        & 1990s          &  0      & 0         & 0         & \dots & 0\\
2        & 1990s          &  0      & 1         & 1         & \dots & 0\\
3        & 1990s          &  0      & 0         & 0         & \dots & 0\\
\dots    & \\
1,682    & 1990s          &  0      & 0         & 0         & \dots & 0\\
\hline
\end{tabular}
\label{subtable: movie}
}
\qquad
\setlength{\tabcolsep}{8pt}
\subtable[Rates]{
\begin{tabular}{cccccccc}
\hline
User-id$\diagdown$ Movie-id &  1     &  2     &  3    & 4      & 5      & \dots & 1,682\\
\hline
1      &  0      &  1      &  0     &  1      & 0     & \dots  & 0\\
2      &  1      &  0      &  0     &  1      & 0     & \dots  & 1\\
3      &  0      &  0      &  0     &  0      & 1     & \dots  & 1\\
\dots  & \\
943    &  0      &  0      &  1     &  1      & 0     & \dots  & 1\\
\hline
\end{tabular}
\label{subtable: rates}
}
\end{table}

\begin{defn}\label{definition: m-m-er}
\cite{MinHuZhu12GranularFour} A many-to-many entity-relationship system (MMER) is a 5-tuple $ES = (U, A, V, B, R)$, where $(U, A)$ and $(V, B)$ are two information systems, and $R \subseteq U \times V$ is a binary relation from $U$ to $V$.
\end{defn}

An example of MMER is given by Table \ref{table: mmer}.

  %
  %
  \subsection{Positive rules}\label{subsection: positive-rules}
Granular association rules reveal patterns in the MMERs.
They connect granules of two universes.
\begin{defn}\label{definition: grarule}
\cite{MinHuZhu12GranularFour} A \emph{granular association rule} is an implication of the form
\begin{equation}\label{equation: granular-association}
(GR): \bigwedge_{a \in A'}\langle a: a(x) \rangle \Rightarrow \bigwedge_{b \in B'}\langle b: b(y) \rangle,
\end{equation}
where $A' \subseteq A$ and $B' \subseteq B$.
\end{defn}

\begin{defn}\label{defition: grarule-positive}
A \emph{granular association rule} indicated by Equation (\ref{equation: granular-association}) is \emph{positive} if both $(A', x)$ and $(B', y)$ are positive granules.
\end{defn}

For brevity, in the following context granular association rules will be simply called \emph{rules}, and positive (negative) granular association rules will be simply called \emph{positive (negative) rules}.
According to Equation (\ref{equation: block-contain-x}), the set of objects meeting the left-hand side of the rule is
\begin{equation}\label{equation: left-granular-rule}
LH(GR) = E_{A'}(x);
\end{equation}
while the set of objects meeting the right-hand side is
\begin{equation}\label{equation: right-granular-rule}
RH(GR) = E_{B'}(y).
\end{equation}

  %
  %
  \section{Positive rule mining}\label{section: rule-mining}
In this section, we first revisit four measures that evaluate the strength of a granular association rule \cite{MinHuZhu12GranularFour}.
Then we define a positive rule mining problem.
The problem is slightly different from the one define in \cite{MinHuZhu12GranularFour} in that it only requires positive rules.
Finally we develop a rule mining algorithm which is similar to the one proposed in \cite{MinZhu12Parametric} for the new problem.

  %
  %
  \subsection{Measures of granular association rule}\label{subsection: four-measures}
From the movielens data set, we may obtain a rule ``35.5\% male students rate 26.7\% thriller movies released in 1990s, 14.4\% users are male students and 12.5\% movies are thriller released in 1990s."
Here 35.5\%, 26.7\%, 14.4\%, and 12.5\% are the source coverage, the target coverage, the source confidence, and the target confidence, respectively.
These measures are defined as follows.
The \emph{source coverage} of a rule is
\begin{equation}\label{equation: source-coverage}
scov(GR) = |LH(GR)| / |U|.
\end{equation}
The \emph{target coverage} of $GR$ is
\begin{equation}\label{equation: target-coverage}
tcov(GR) = |RH(GR)| / |V|.
\end{equation}

There is a tradeoff between the source confidence and the target confidence of a rule.
Consequently, neither value can be obtained directly from the rule.
To compute any one of them, we should specify the threshold of the other.
Let $tc$ be the target confidence threshold.
The \emph{source confidence} of the rule is
\begin{equation}\label{equation: source-confidence}
sconf(GR, tc)
= \frac{|\{x \in LH(GR) | \frac{|R(x) \cap RH(GR)|}{|RH(GR)|} \geq tc\}|}{|LH(GR)|}.
\end{equation}

  %
  %
  \subsection{The positive rule mining problem}\label{subsection: problem}
Now we propose a rule mining problem as follow.
\begin{problem}\label{problem: rule-mining}
The positive rule mining problem.

\textbf{Input:} An $ES = (U, A, V, B, R)$, a minimal source coverage threshold $ms$, a minimal target coverage threshold $mt$, a minimal source confidence threshold $sc$, and a minimal target confidence threshold $tc$.

\textbf{Output:} All positive rules satisfying $scov(GR) \geq ms$, $tcov(GR) \geq mt$, $sconf(GR) \geq sc$, and $tconf(GR) \geq tc$.
\end{problem}

This problem is quite similar to the one discussed in \cite{MinHuZhu12GranularTwo,MinHuZhu12GranularFour,MinZhu12Parametric,MinZhu12ParametricCold}.
The only difference is that positive rules instead of all rules are output.

  %
  %
  \subsection{A backward algorithm}\label{section: algorithm}
\begin{algorithm}[tb!]\caption{A backward algorithm}\label{algorithm: partial-rule-mining-backward}
  \textbf{Input}: $ES = (U, A, V, B, R)$, $ms$, $mt$, $sc$, $tc$.\\
  \textbf{Output}: All positive rules satisfying given thresholds.\\
  \textbf{Method}: backward\\
  \begin{algorithmic}[1]
    \STATE $SG(ms) = \{(A', x) \in 2^A \times U | (A', x)$ is positive, $\frac{|E_{A'}(x)|}{|U|} \geq ms\}$;
    \STATE $TG(mt) = \{(B', y) \in 2^B \times V | (B', y)$ is positive, $\frac{|E_{B'}(y)|}{|V|} \geq mt\}$;
    \FOR {each $g' \in TG(mt)$}
      \STATE $Y = e(g')$;
      \STATE $X = \underline{R^{-1}}_{tc}(Y)$;
      \FOR {each $g \in SG(ms)$}
        \IF {($|X \cap e(g)| / |e(g)| \geq sc$)}
          \STATE output rule $i(g) \Rightarrow i(g')$;
        \ENDIF
      \ENDFOR
    \ENDFOR
  \end{algorithmic}
\end{algorithm}

We propose an algorithm to deal with Problem \ref{problem: rule-mining}.
The algorithm is listed in Algorithm \ref{algorithm: partial-rule-mining-backward}.
It is very similar to the algorithm proposed in \cite{MinZhu12ParametricCold}.
The difference lies in the first two lines.
To implement these lines, we first produce positive granules with length one.
For example, ``gender is male," ``genre is thriller" and ``release decade is 1990s".
Then we follow the structure of the Apriori algorithm to produce longer positive granules.
For example, ``genre is thriller and release decade is 1990s", or equivalently, ``thriller movies released in 1990s".
In this way, only positive granules are generated.
The conditions $\frac{|E_{A'}(x)|}{|U|} \geq ms$ and $\frac{|E_{B'}(y)|}{|V|} \geq mt$ ensure that only positive granules satisfying coverage measures are kept.

Lines 3 through 10 mine rules satisfying confidence measures.
To explain these codes we should revisit the definition of lower approximation on two universes.
\begin{defn}\label{definition: parametric-lower-approximation}
\cite{MinZhu12ParametricCold} Let $U$ and $V$ be two universes, $R \subseteq U \times V$ be a binary relation, and $0 < \beta \leq 1$ be a user-specified threshold.
The lower approximation of $X \subseteq U$ with respect to $R$ for threshold $\beta$ is
\begin{equation}\label{equation: variable-forward-neighbor-lower-approximation}
\underline{R}_{\beta}(X) = \{y \in V | \frac{|R^{-1}(y) \cap X|}{|X|} \geq \beta\}.
\end{equation}
\end{defn}

From this definition we know immediately that the lower approximation of $Y \subseteq V$ with respect to $R$ is
\begin{equation}\label{equation: parametric-backward-neighbor-lower-approximation}
\underline{R^{-1}}_{\beta}(Y) = \{x \in U | \frac{|R(x) \cap Y|}{|Y|} \geq \beta\}.
\end{equation}
Here $\beta$ corresponds with the target confidence instead.
The lower approximation can help speeding up the mining process.
This issue has been discussed in \cite{MinZhu12ParametricCold}, and similar phenomenon holds for our problem.

  %
  %
  \section{Experiments on the movielens data set}\label{section: experiments}
The Internet Movie Database \cite{movielens} is widely used in recommender systems (see, e.g., \cite{ScheinA2002ColdStart}).
It contains 100,000 ratings (1-5) from 943 users on 1,682 movies, with each user rating at least 20 movies.
The main purpose of our experiments is to answer the following questions.
\begin{enumerate}
\item{Does the priority-based approach lose important information?}
\item{Is it necessary to remove negative rules?}
\end{enumerate}

  %
  %
  \subsection{Priority-based approach vs. scaling-based approach}\label{subsection: priority-scale}
With the priority-based approach, we assign a priority to each genre and keep the most important one for a movie \cite{MinZhu12Parametric}.
In this way, no negative rule exists.
However, some information is lost.
In contrast, with the scaling-based approach, no information is lost, and negative rules are filtered out by Algorithm \ref{algorithm: partial-rule-mining-backward}.
Now we compare the number of positive rules that are generated through these two approaches.
We use the following setting:\\
(Setting 1) $sc = tc = 0.1$,  $ms = mt$,  and $mt \in [0.05, 0.12]$.

\begin{figure}[tb]
    \begin{center}
    \includegraphics[width=2.7in]{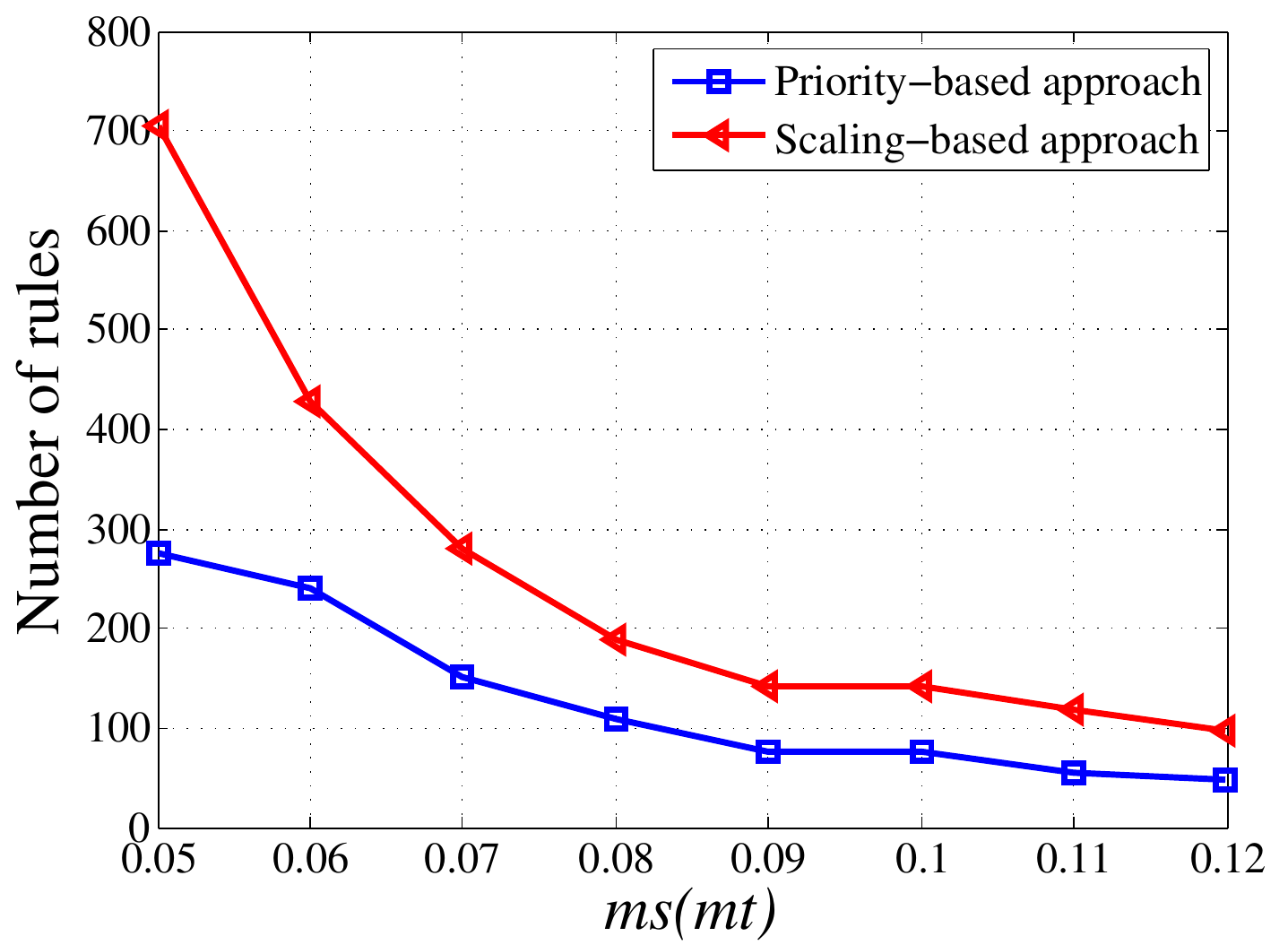}
    \caption{Rules mined through priority-based and scaling-based approaches}
    \label{figure: priori-vs-scale}
    \end{center}
\end{figure}

Results are depicted in Figure \ref{figure: priori-vs-scale}.
Here we observe that the number of rules mined through the scaling-based approach is more than twice of the priori-based approach.
Therefore the information lost by the priority-based approach is unacceptable in applications.

  %
  %
  \subsection{Influence of negative rules}\label{subsection: influence-negative}
In applications, we want the recommender system to generate a number of rules.
This number should not be too big; otherwise it will be impossible for users to pick up interesting and useful ones.
Therefore we need to specify thresholds on four measures carefully such that a few to a few hundred rules are generated.

\begin{figure}[tb]
    \begin{center}
            \subfigure[]{
            \begin{minipage}[b]{2.5in}
             \includegraphics[width=2.45in]{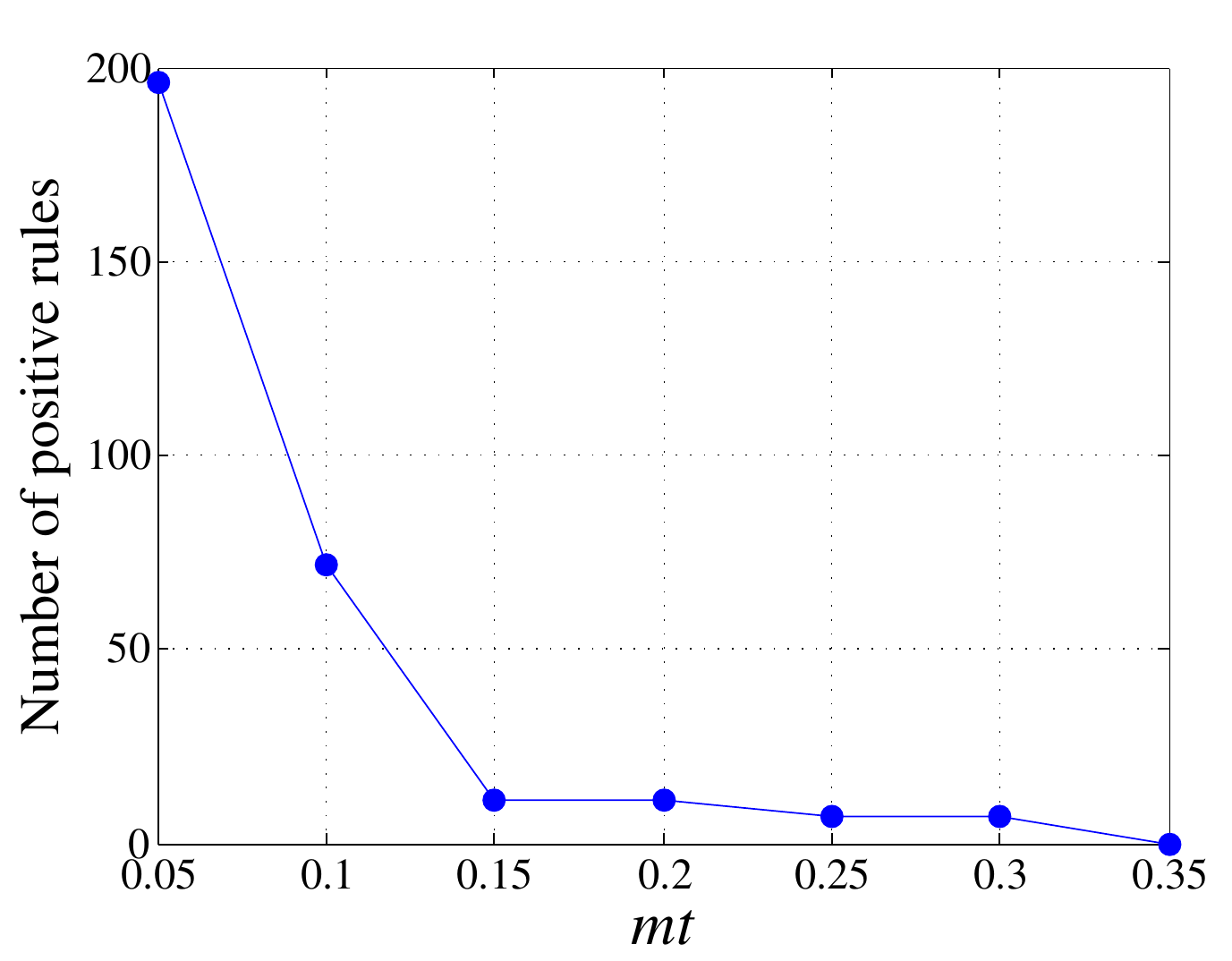}
            \end{minipage}
            }
            \subfigure[]{
            \begin{minipage}[b]{2.5in}
             \includegraphics[width=2.45in]{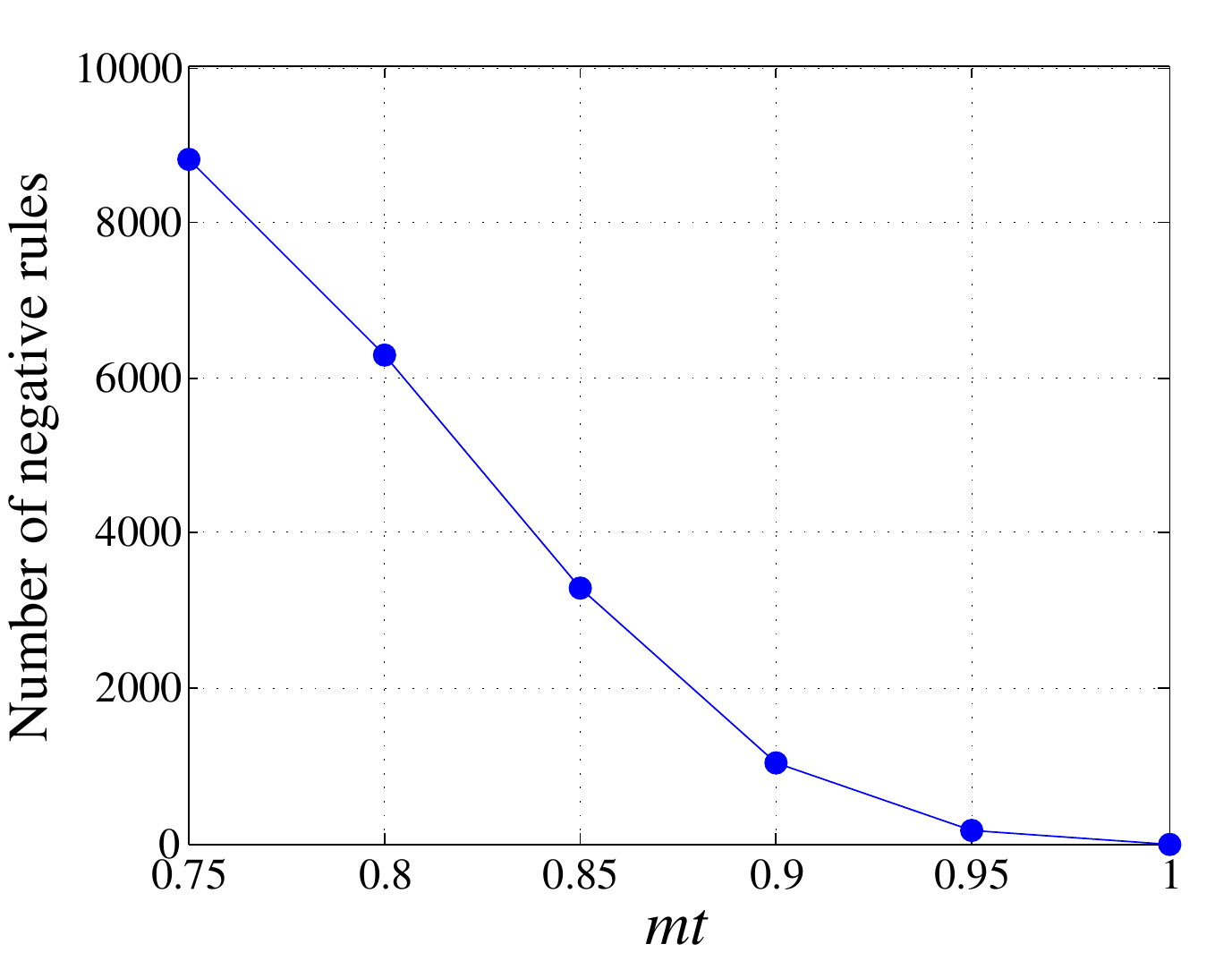}
            \end{minipage}
            }
\end{center}
            \caption{$sc = 0.12$, $tc = 0.15$, $ms = 0.1$ (a) Number of positive rules; (b) Number of negative rules.}
            \label{figure: neg-pos-rules-trend}
\end{figure}

First, we generate rules using the algorithm presented in \cite{MinZhu12ParametricCold}.
We use the following setting: \\
(Setting 2) $ms = 0.1$, $mt = 0.85$, $sc = 0.12$, $tc = 0.15$.\\
With this setting we obtain 3,300 rules.
Some of them are given below:

\noindent(Rule 1) $\langle\textrm{Gender, Male}\rangle \wedge \langle \textrm{Occupation, Student} \rangle(136)$ \\
\indent $\Rightarrow \langle \textrm{Adventure, 0}\rangle\wedge \langle \textrm{Mystery, 0}\rangle(1488)$\\
$[scov = 0.144, tcov = 0.884, sconf = 0.132, tconf = 0.150]$

(Rule 2) $\langle\textrm{Gender, Male}\rangle \wedge \langle \textrm{Age, (0, 18]} \rangle(136)$\\
\indent$\Rightarrow \langle \textrm{Animation, 0}\rangle \wedge \langle \textrm{War, 0}\rangle(1569)$\\
$[scov = 0.144, tcov = 0.933, sconf = 0.132, tconf = 0.150]$

The target coverage threshold is $mt = 0.85$, therefore these rules are very strong from this viewpoint.
Unfortunately, they are all negative rules, and they are not quite interesting.
Rule 1 is read as ``Male students rate movies that are \emph{neither} Adventure \emph{nor} Mystery, 136 users are male students and 1,488 movies are \emph{neither} Adventure \emph{nor} Mystery."
It is straight forward to compute the source/target coverage.
The source coverage of the rule is $136/943 \approx 0.1442 > 0.1$, and the target coverage is $1488/1682 \approx 0.8847 > 0.85$.
As discussed earlier, we cannot obtain the source/target confidence directly.
We only know that they exceed the given thresholds.

Second, we generate positive rules using Algorithm \ref{algorithm: partial-rule-mining-backward}.
We use the following setting:\\
(Setting 3) $ms = 0.1$, $mt = 0.1$, $sc = 0.12$, $tc = 0.15$.\\
This setting is different from Setting 2 only on $mt = 0.1$.
With this setting we obtain 72 positive rules.
Some of them are given below:\newline
(Rule 3) $\langle\textrm{Gender, Male}\rangle \wedge \langle \textrm{Age, (0, 18]} \rangle(136)$\\
\indent$\Rightarrow \langle \textrm{Year, 1990s}\rangle\wedge \langle \textrm{Action, 1}\rangle(206)$\\
$[scov = 0.144, tcov = 0.122, sconf = 0.250, tconf = 0.150]$

(Rule 4) $\langle\textrm{Gender, Male}\rangle \wedge \langle \textrm{Occupation, Student} \rangle(136)$\\
\indent$\Rightarrow \langle \textrm{Year, 1990s}\rangle\wedge \langle \textrm{Thriller, 1}\rangle(211)$\\
$[scov = 0.144, tcov = 0.125, sconf = 0.220, tconf = 0.150]$

Rule 3 is read as ``Young men no more than 18 years old rate action movies released in 1990s."
This is quite interesting to us even though there are only 206 action movies released in 1990s.

Third, we observe the change of the number of rules with different settings of $mt$.
Figure \ref{figure: neg-pos-rules-trend}(a) shows the number of rules for $mt \in [0.75, 1]$.
Unfortunately, all rules are negative ones.
To produce positive rules, we should use lower $mt$.
Figure \ref{figure: neg-pos-rules-trend}(b) shows the number of positive rules for $mt \in [0.05, 0.35]$.
We observe that there would be no positive at all for $mt \geq 0.35$.
However, according to Figure \ref{figure: neg-pos-rules-trend}(a), there are about 9,000 negative rules for $mt = 0.75$.
Hence we cannot generate negative rules for $mt \geq 0.35$ since there are too many of them.
In other words, if we do not filter out negative granular association rules, they will overwhelm positive ones.

  %
  %
  \subsection{Discussions}\label{subsection: discussions}
Now we can answer the questions proposed at the beginning of this section.
\begin{enumerate}
\item{The priority-based approach loses important information and rules.
The scaling-based approach, on the other hand, keeps all information and helps mining all positive rules.}
\item{It is very important to remove negative rules because they overwhelm positive ones.}
\end{enumerate}

  %
  %
  \section{Conclusions and further works}\label{section: conclusion}
In this paper, we deal with multi-value data with two objectives.
The first is to keep all useful information such that all interesting granular association rules can be mined.
This is achieved through attribute scaling.
The second is to remove strong however uninteresting rules.
This is achieved through filtering out negative granules and negative rules.
In the future, we will address other types of data such as interval valued for more applications.

  %
  %
  \section*{Acknowledgements}\label{section: acknowledgements}
This work is in part supported by National Science Foundation of China under Grant No. 61170128, the Natural Science Foundation of Fujian Province, China under Grant Nos. 2011J01374, 2012J01294, and Fujian Province Foundation of Higher Education under Grant No. JK2012028.

  %
  %

\end{document}